\documentclass[apj]{emulateapj}
\usepackage{xspace}
\journalinfo{Accepted for publication in the Astrophysical Journal}
\slugcomment{To appear in ApJ v601 n2, 1 Feb 2004}

\newcommand\msun{\ensuremath{M_\sun}\xspace}
\newcommand\mcrit{\ensuremath{M_{crit}}\xspace}

\newcommand\ubv{\ensuremath{U\!BV}}
\newcommand\ebv{\ensuremath{E(\bv)}\xspace}

\shorttitle{White dwarfs in binaries}
\shortauthors{Williams}

\begin{document}

\title{The impact of unresolved binaries on searches for white dwarfs
  in open clusters} 

\author{Kurtis A. Williams}
\affil{Steward Observatory}
\affil{933 N. Cherry Ave., Tucson, AZ 85721}
\email{kurtis@as.arizona.edu}

\begin{abstract}
Many open clusters have a deficit of observed white dwarfs (WDs) compared
with predictions of the number of stars to have evolved into WDs.  We
evaluate the number of WDs produced in open clusters and the number
of those WDS detectable using photometric selection techniques. This
calculation includes the effects of varying the initial-mass function (IMF), 
the maximum progenitor masses
of WDs, and the binary fraction.  Differences between the
calculated number of observable WDs and the actual number of WDs
observed in a specific cluster then indicate the true deficit of WDs
that must be explained through effects such as dynamical evolution of
the cluster or close binary evolution.
Observations of WDs in three open clusters, the Hyades,
Pleiades, and Praesepe, are compared to the calculated observable
populations in those clusters. The
results suggest that a large portion of the white dwarf deficit may be
explained by the presence of WDs in unresolved binary systems. However,
the calculated WD populations  still over-predict the number of
observable WDs in each cluster.  While these calculations cannot
determine the cause of this residual white dwarf deficit, potential
explanations include a steep high-mass IMF, dynamical evolution of the cluster,
or an increased likelihood of equal-mass components in a binary
system.  Observations of complete WD samples in open clusters covering
a range of ages and mass can help to distinguish between these possibilities.
\end{abstract}


\section{Introduction}
White dwarfs (WDs) are the final endpoint of stellar evolution for the vast
majority of stars.  WDs are gaining importance in tracing the history
of stellar populations in the galaxy, including the galactic disk
\citep[e.g.][]{Winget87b}, open star clusters
\citep[e.g.][]{Richer98}, and globular star clusters
\citep{Hansen02}.  WDs are also useful in studies of supernova
physics, as the upper mass of WD progenitors represents the
critical progenitor mass (\mcrit)\footnote{Also
  referred to as $M_w$, $m_w$, and $M_{up}$} for core-collapse
supernovae. The value of \mcrit is relatively
uncertain, with the best estimates of $5.5\msun\lesssim\mcrit\lesssim 10\msun$.  

The first significant population of white dwarfs (WDs) in open clusters was
identified by \citet{Eggen65} in the \objectname{Hyades},
\objectname{Praesepe}, and the \objectname{Pleiades}.
\citet{Tinsley74} compared the number of Hyades WDs to the estimated
number of Hyades stars having completed their lifetimes. Her work
found that the WD numbers are compatible with $\mcrit\sim 4\msun$,
though this determination depends sensitively on the assumed
initial-mass function (IMF), the turnoff mass of the Hyades, and the Hyades distance
modulus. This value of \mcrit is lower than current lower limits
($\mcrit\gtrsim 6\msun$), set in large part by the existence of \objectname{LB 1497},
the lone WD in the \objectname{Pleiades} (turnoff mass
$\sim 5.4\msun$).

\citet{Weidemann92} discuss the Hyades WD population and determine
a deficit of 21 WDs (28 WDs predicted versus 7 observed WDs) for
$\mcrit=8\msun$ and ascribe this white dwarf deficit to dynamical
evaporation of WDs from the Hyades.  Similar deficits of WDs have
since been observed in other open clusters, including \objectname{M67}
\citep{Richer98}, \objectname{Praesepe} (Claver et al. 2001,
hereafter \citet{Claver01}), and \objectname{NGC 2099} \citep{Kalirai01b}.  The existence of the white dwarf deficit
is still debatable, as discussed by \citet{vonhippel98}.

Three explanations for the white dwarf deficit in open clusters
readily come to mind.  
First, WDs may evaporate from open clusters due to dynamical
evolution.  Mass segregation and galactic tidal fields alone are not
sufficient to remove WDs from open clusters, as a 0.6\msun WD is still
more massive than typical cluster stars and thus unlikely to suffer
preferential evaporation.  This is borne out by modern $N$-body
simulations of open clusters, such as those of
\citet{Zwart01,Baumgardt03} and \citet{Hurley03}, which find that WDs
remain bound in open clusters.  However, if a WD receives a sufficient
velocity kick from asymmetric mass loss during its post-main sequence
evolution, the WD may become unbound from the open cluster.  This
scenario was first suggested by  \citet{Weidemann92} to explain the WD
deficit in the \objectname{Hyades}. Recent, simple $N$-body
simulations by \citet{Fellhauer03} confirm that this mechanism can
preferentially remove WDs from an open cluster, though scant
observational evidence exists.

A second explanation for the white dwarf deficit, though not exclusive
of the dynamical evolutionary argument, is that the WDs may
be hidden in binary systems in which the intrinsically faint WDs
are not detected due to the overwhelming light of brighter companion
\citep[e.g.][]{Kalirai01b}.  
Searches for WDs in binaries have been undertaken in the
Hyades. Two of the known Hyades WDs,
\objectname{EGGR 38} and \objectname{V471 Tau}, are in
binary systems. More recently, additional Hyades WDs have been
discovered hidden in unresolved binaries, including \objectname{HD 27483}
\citep{Bohmvitense93}, \objectname{VA351} \citep{Franz98}, and 
four potential WDs in Am binary star systems
\citep{Debernardi00}. Clearly some WDs lie hidden in unresolved
binary systems, but the exact numbers are not known. 

The third explanation for the small number of white dwarfs seen
in the Hyades and other well studied young clusters could be
that our expectations of WD numbers are incorrect. 
The deficit discussed for the
Hyades would go away if \mcrit is low ($\sim
4\msun$), if the IMF for masses above the cluster turnoff mass is
steeper than the present-day mass function around the turnoff mass,
or a combination of these two effects.
The low value of \mcrit seems unlikely, given that WDs have been
observed in clusters with turnoff masses higher than 4\msun, including
the Pleiades, \objectname{NGC 2516} \citep{Koester96} and \objectname{NGC 2168}
\citep{Reimers88}. 

We have developed a Monte Carlo method of calculating the number of
WDs detectable in observations of specific open clusters, a
calculation designed to aid in the interpretation of data from
the Lick-Arizona White Dwarf Survey \citep[LAWDS, ][]{Williams03} and
could be used in conjunction with other ongoing cluster WD surveys,
such as the CFHT Open Cluster Survey \citep{Kalirai01} and the WIYN
Open Cluster Study \citep{vonhippel98b}. 
This calculation utilizes the observed characteristics of specific
open clusters, including the cluster age, distance and reddening, and
determines how many WDs would be detected 
given photometric WD selection criteria.  The structure of this
calculation, its limitations, and tests of the calculation
 are presented in \S2.
In \S3 we present the results of the calculations
for
the Hyades, Praesepe, and the Pleiades and compare these results to
the observed WD populations.  In \S4 we discuss
the results and discuss the usefulness of these calculations in regard
to current searches for WDs in open clusters.  

\section{Calculating the WD populations in open clusters \label{section.sim}}
The number of WDs detectable by photometric observations of open
clusters can be calculated by appropriate modeling of the cluster,
including the effects of unresolved binary stars, choice
of IMF, WD detection criteria, \mcrit, metallicity and cluster age.  
Assuming that such an open cluster model
is sufficiently realistic, any significant difference
between the calculated numbers of observable WDs and the actual numbers
of observed WDs represents the ``true'' white dwarf deficit that must be
explained by other means, such as evaporation or other dynamical evolution.

Monte Carlo techniques are used to calculate the observable WD
population. Stars are randomly drawn from an
input IMF. For this work, four input IMFs $\xi(M)\propto
M^{-(1+\Gamma)}$ were considered:
a Salpeter IMF  with $\Gamma=1.35$ \citep{Salpeter55}, a steeper power
law that accounts for unresolved binary systems with $\Gamma=2$
\citep{Naylor02}, and a broken power law with a flat low-mass slope:
$\Gamma=0.2\,(M\leq 1\msun)\, ; \, \Gamma=1.8\,(M> 1\msun)$
\citep{Naylor02}, hereafter called the ``Naylor IMF,'' and the Kroupa
IMF with  $\Gamma=0.3\,(M<0.5\msun)\, ; \, \Gamma=1.3\,(M\geq
0.5\msun)$ \citet{Kroupa01}. 

Stars are assigned a binary
companion with a probability based on an input binary fraction; the
binary mass fraction is assumed to be random, i.e. the primary and
secondary stars are drawn from the same IMF (see \S\ref{limitations} below).  
The stars are then evolved to the input cluster age using the $Z=0.008,
Z=0.019$ or $Z=0.040$ solar-scaled metallicity stellar
evolutionary models of \citet{Salasnich00}.  These models provide
evolutionary data on stars with masses up to 20\msun, sufficiently
massive to explore all likely values of \mcrit.

Two outcomes are assumed
for stars having completed their evolution. 
Stars  with $M_* > \mcrit $ undergo supernovae.  Stars
with $M_* \leq \mcrit$ form a WD.  The WD
mass is determined from a linear approximation of the initial-final
mass relation in \citet{Weidemann00},
\begin{equation}
M_{WD} = \frac{3}{35}M_{*} + 0.414 \msun \, .
\end{equation}
Each WD is assigned a cooling age equivalent to the progenitor
lifetime subtracted from the input cluster age. The radii, effective
temperature, and surface gravities of each WD are interpolated from
the carbon/oxygen models of Wood (1995, hereafter \citet{Wood95}).

Photometric indices for each system are determined by summation of the flux
from each component. Supernova remnants are assumed to emit no
flux. WD fluxes are determined from the updated model atmospheres
of \citet{Finley97} graciously provided by D. Koester.  
Stellar fluxes are taken from the
isochrones of \citet{Girardi02} for the \citet{Salasnich00}
evolutionary models.  These indices are corrected for the input
distance modulus and reddening, assuming the interstellar reddening
law of \citet{Rieke85} and $R_V=3.1$.

Once the photometric indices for each system have been calculated,
these indices are compared to the input WD photometric-selection 
criteria. These criteria include color indices and single-band
magnitude limits, with the intention that the photometric criteria
used to select WD candidates in actual observations of a given cluster
are used in the simulation of that cluster.
If a system meets these input photometric selection
criteria, the number of detected WDs is incremented
by one.  If it fails to meet the selection criteria, the WD is
considered undetected. 

The Monte Carlo calculation continues to select stellar systems until
the number of stars brighter than a limiting magnitude $V_{lim}$ reaches  the
input value.  For example, if the cumulative luminosity function of an
actual open cluster has $N_*$ stars brighter than $V_{lim}$, then each
simulated cluster will also contain $N_*$ stars brighter than
$V_{lim}$.  $10^4$ realization of each cluster are used to ensure
high precision in the output.

\subsection{Limitations of the calculations\label{limitations}}

These calculations represent a  simplified version of actual open
clusters.  The simulation assumes that all binary stars are
unresolved, an assumption whose validity depends strongly on the
cluster distance and the distribution of physical separations among
binary stars.  Binary interactions during stellar
evolution, such as mass transfer systems, common envelope stages,
cataclysmic variables, and the potential 
disruption of binaries due to mass loss
are ignored. Open cluster dynamical evolution, such as
evaporation of stars and evolution of the binary fraction, is also
ignored.  The binary mass ratios are assumed to be random, which is likely
not be the case for close binaries, but is reasonable for wider
binaries \citep{Mason98,Abt90}. 
Finally, stellar systems with more than two components are not
considered. 

Ignoring binary-star evolution and the different mass ratios of close
binaries than a random IMF results in far fewer double-degenerate
systems being created than as seen in realistic $N$-body simulations,
such as the significant double-degenerate population seen in
\cite{Hurley03}.  The presence of any observed double-degenerate
sequence in an open cluster therefore can provide a constraint on the
relative importance of close-binary evolution in open clusters.

The lack of consideration of dynamical evolution is not
a deficiency in the calculations, but rather is one purpose of
these calculations -- to determine the actual number of missing WDs that
dynamical evolution must be invoked to explain.  The strong, observed
evaporation of low-mass stars has no great 
impact on this calculation.  Low-mass stars in binaries with a
(comparatively) high-mass WD should not be affected greatly by
evaporation processes, and low-mass stars not in a binary system with
a WD are not used in the calculation beyond the initial selection of
cluster stars.

In short, these calculations are not a
simulation of WDs in the larger sense of open cluster evolution.  
Rather, this calculation provides a
more sophisticated means of comparing the observed WD populations of
specific open clusters to the expected observable WD population.  Any
differences between the actual observed WD population and the
calculated observable population provide a measure of the strength of
the dynamical evolution of the WD population.

\subsection{Trials of the calculations\label{section.test}}
\begin{figure}
\plotone{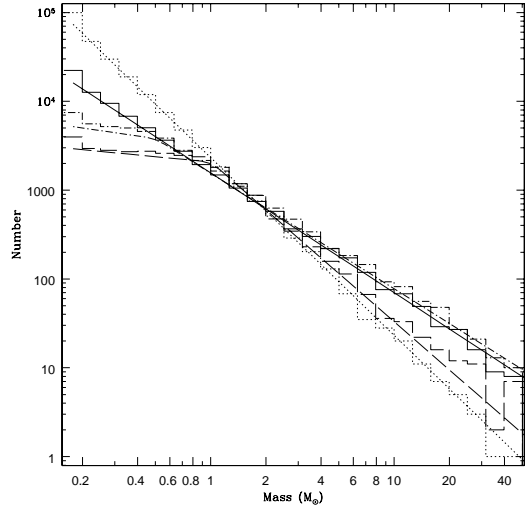}
\caption{Simulated initial-mass functions.  Histograms indicate the
  output for a Salpeter IMF (solid), a $\Gamma=2$ power-law IMF
  (dotted), Naylor IMF (long-dashed) and Kroupa IMF (dash-dotted).  Lines
  indicate the expected number of stars for each
  IMF. \label{fig.plot_imf}}
\end{figure}

We ran numerous tests to verify that these calculations return reasonable
results for certain limiting cases.  
Fig.~\ref{fig.plot_imf} shows the output IMF
of a single simulated cluster with 5000 stars more massive
than 1\msun for each of the input IMFs.  As one would hope, the output
IMFs agree with the input IMFs.  

Simulated clusters with a $\Gamma=2$ power-law or the Naylor
IMF are found to contain fewer WDs than those with a Salpeter or
Kroupa IMF,
merely due to the fact that fewer high-mass stars formed due to the
steeper slope.  The Naylor and Kroupa IMFs result in a
higher fraction of WDs being hidden in binary systems than the single
power-law
IMFs due to
the relatively flat low-mass IMF slope, as a companion to a WD is more
likely to be higher mass and therefore more luminous than the WD.  

\begin{figure}
\plotone{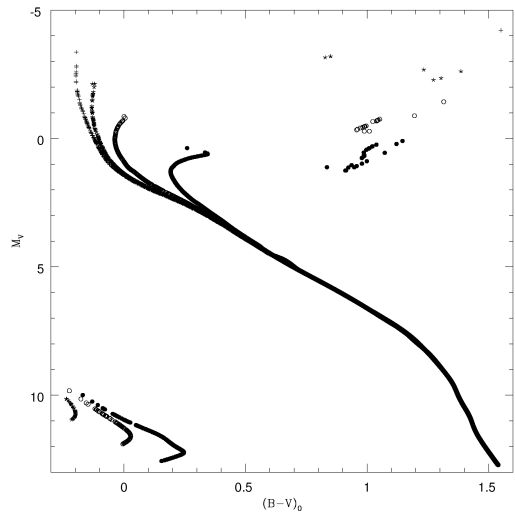}
\caption{$\bv,\, M_V$ CMD for simulated clusters of various ages.
 Simulated ages are for logarithmic cluster ages of 7.5 (pluses), 8.0
 (asterisks), 8.5 (open circles), and 9.0 (filled squares).
 Simulated clusters have 1000 existing stars more massive than
 1\msun. \label{fig.plot_ages}} 
\end{figure}

Fig.~\ref{fig.plot_ages} shows output color-magnitude diagrams (CMDs)
for calculations of clusters at four different ages containing no binary
stars.  Occasional minor numerical issues arise
during the interpolation of photometric indices  for evolved stars 
due to the large changes in magnitudes with minute changes in stellar
mass.  These issues arise primarily at inflection points in the
isochrones, such at the helium flash and at the tip of the AGB.
These errors do not affect the simulated detection of WDs, as
both the correct and erroneous magnitudes for these stars are
bright enough to hide any WD companions, and the number of stars
affected is very small.

\begin{figure*}
\plottwo{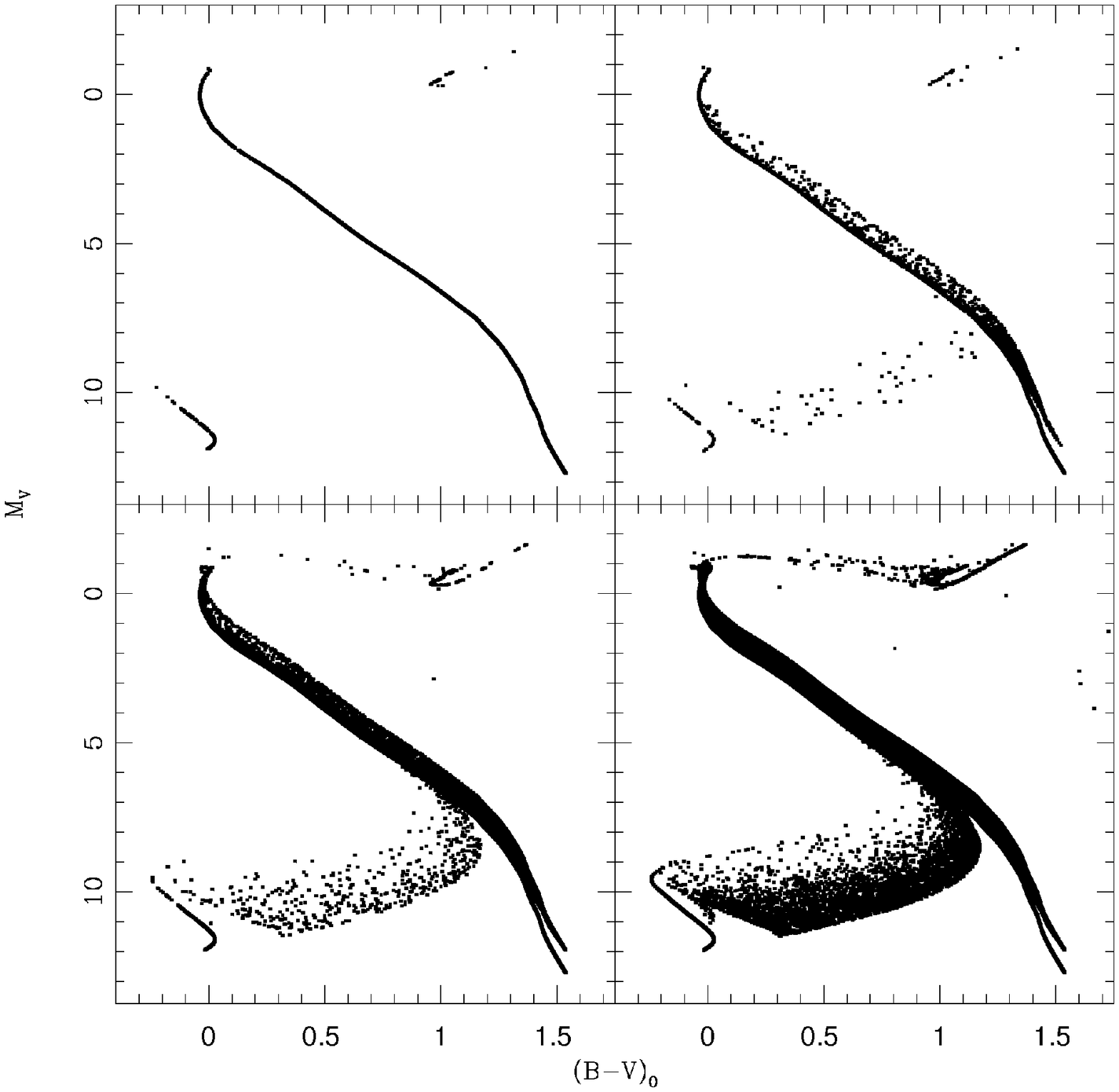}{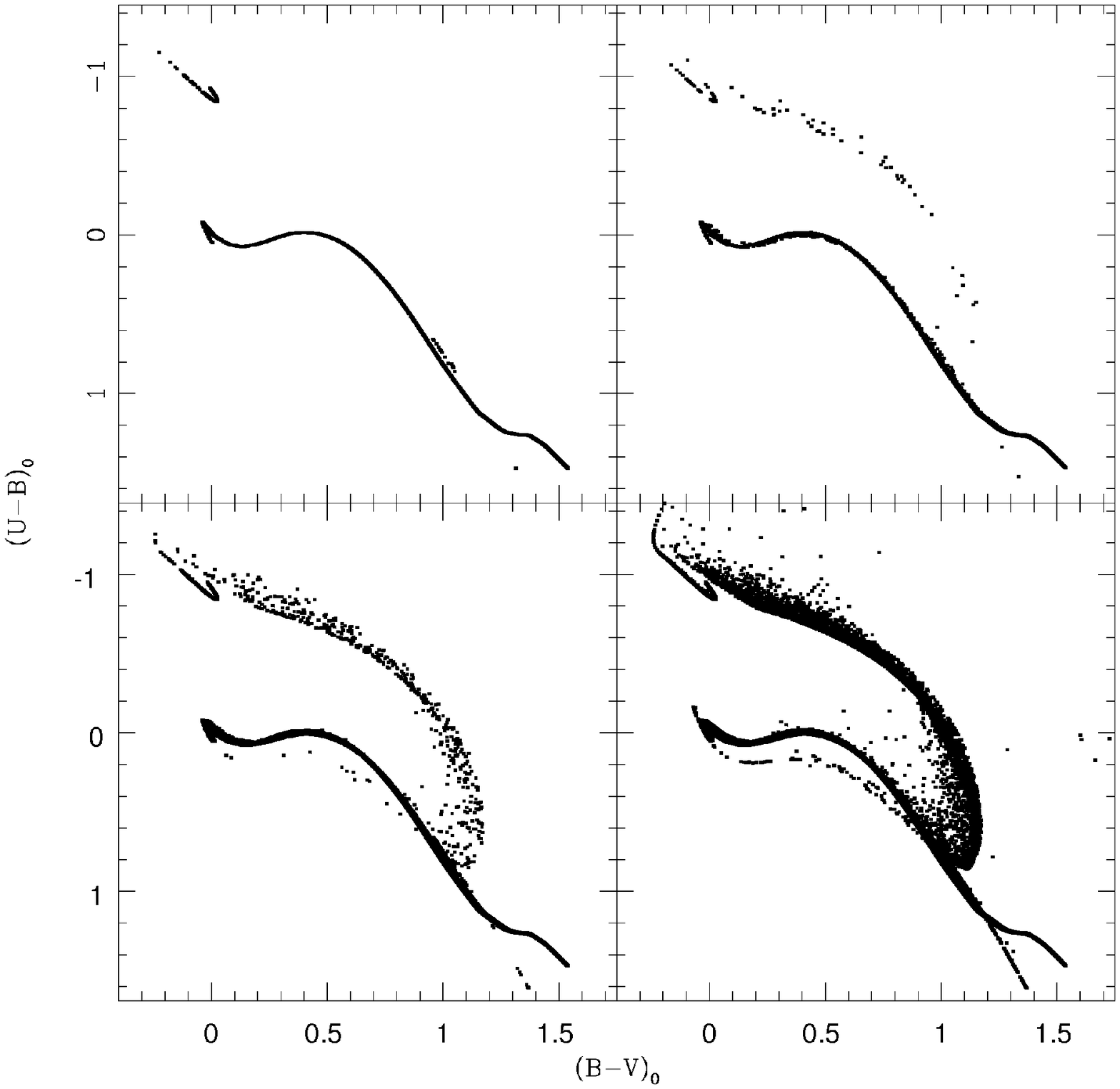}
\caption{$\bv,\,M_V$ CMD (left) and $\bv,\,\ub$ color-color diagram
  (right) for simulated clusters with binary stars.  For each diagram,
  the number of stars more massive than 1\msun is 1000 (top panels),
  $10^4$ (bottom left) and $10^5$ (bottom right). The top left panel
  in each diagram is for a simulated cluster with no binary stars; the
  binary fraction for each of the other panels is 0.5.  The WD-WD
  binary sequence is visible in the lower right panel of the CMD, 0.75
  mags brighter than the lone WD cooling sequence. Each simulation
  is of a cluster with a logarithmic age of 8.50 and the broken
  power-law IMF.\label{fig.plot_bin}}
\end{figure*}

The effect of the addition of binary stars is shown in
Fig.~\ref{fig.plot_bin}.  The
sequence of binaries consisting of two main-sequence (MS) stars is readily
visible in the $\bv, M_V$ CMD.  The fan of points leading from the
MS to the WD cooling sequence is composed of MS-WD binaries. The
$\bv,\ub$ color-color diagram illustrates the advantage of three-band
photometry over two-band photometry.  
The WD-MS binaries, which would blend in with background disk MS stars
in a $\bv, V$ CMD, are well-separated from the locus of main-sequence
in color-color space.  In addition, the color-color space information
clearly separates the WD cooling sequence from massive MS stars.

A WD-WD
binary sequence is visible only when the number of stars in a
cluster is very high, as seen in the
lower-right panel of CMD in Fig.~\ref{fig.plot_bin}.  Any significant
observed 
WD-WD binary population in open clusters would therefore
confirm that the binary component mass ratios are not random
and/or that dynamical processes such as binary interactions are not
negligible.  \citet{Hurley03} find evidence of a WD-WD sequence in the
photometry of NGC 6819 of \citet{Kalirai01c}.

Multiple ($10^4$) realizations of each cluster are used to calculate
precise means and standard deviations for various categories of WDs,
including the total number of WDs, the total observed number of WDs,
the number of WDs in binary systems, etc.  

\section{Simulations of specific clusters\label{section.clusters}}

We have calculated the predicted observed WD populations for  three
open clusters: the
Hyades, the Praesepe, and the Pleiades. The output populations
are compared to the number of observed WDs in each
system.   The calculated numbers of
WDs in each cluster fulfilling various criteria, including
whether the WD met the photometric selection criteria (these are given
individually for each cluster below), whether a WD was in a
binary system, and WD progenitor masses were calculated, and the
mean and standard deviation of each over the ensemble of trials 
were calculated.  

The standard
deviations are found to be nearly equal to those expected from Poisson
statistics.  Therefore, we use Poisson statistics to compare the
calculated populations to observations of the actual cluster WDs.  The
probability $P$ of obtaining the actual number of observed WDs in a
category or fewer is calculated from Poisson statistics given the mean value
obtained from the Monte Carlo calculation.  If  $P \leq 0.05$, then 
the derived WD population is considered to be inconsistent with the 
observed population.
For probabilities $0.15\geq P > 0.05$,  the calculated is
termed to be mildly inconsistent with the data.

\emph{The Hyades} --- The \objectname{Hyades} is the open cluster with the
best-studied WD system. Nine WDs are traditionally considered cluster
members, though some questions as to the membership of some of these
have been raised \citep{Weidemann92}. Of these, the seven lone WDs and
the WD-MS binary \objectname{EGGR 38} are detectable as blue excess
objects in \ubv photometry. 

For the Monte Carlo realizations the Hyades, the following physical
parameters are used.  The intrinsic distance modulus is assumed to be 3.33
\citep{Perryman98}, the color excess $E(\bv)=0.01$ \citep{WEBDA}, and
the logarithmic age is assumed to be 8.80 \citep{Perryman98}.  The
binary star fraction is chosen to be 0.4, approximately  the
lower limit of $0.41\pm0.05$ determined from the combination of
speckle, spectroscopic, and direct imaging surveys
\citep{Patience98}.  

The Hyades WDs have been detected and confirmed through a variety of
photometric, spectroscopic, and astrometric surveys, and a
direct comparison of these selection criteria with the calculated
population is not performed.
In order to determine if a cluster WD is to be considered observable,
we apply the
photometric criteria of $\ub\leq 0.0$ and $\bv\leq 0.6$ to select
WDs. These values encompass the range of indices 
observed in the WD sample of
\citet{Eggen65}, including the seven single Hyades WDs and the binary
\objectname{EGGR 38}.  The MS-WD binaries not selected via these
criteria represent the contribution of binaries to the WD deficit.

The metallicity of the Hyades is super-solar, with 
$[\mathrm{Fe}/\mathrm{H}]=0.14\pm 0.05$ \citep{Perryman98}, corresponding
to $Z=0.034$ for a solar $Z=0.0245$ \citep{GONG}.  We therefore
calculate WD populations for each of 
two metallicities, $Z=0.019$ and $Z=0.040$,
for comparison. The
calculated population also explores three values of \mcrit,
10\msun, 8\msun, and
6\msun, bracketing the range of likely values.  Finally, the
calculations were performed using each of the four IMFs discussed above.

Each realization of the cluster contains 138
stellar systems in the magnitude range
$3.0\leq B\leq 10.0$, consistent with the
Hyades luminosity function of \citet{Oort79}, a complete luminosity
function of the Hyades upper main sequence.
The resulting WD populations of the Hyades are given in Table
\ref{tab.hyades_sim}. 

Our results indicate that the eight photometrically-selected
Hyades WDs are inconsistent ($P\leq 0.03$) with what would be expected
from a Salpeter or Kroupa IMF, independent of the value of \mcrit.  However,
the calculated number of WD detections for the $\Gamma=2$ power-law IMF and the
Naylor IMF are fully consistent with the observed number of
WDs for $Z=0.019$, and consistent ($P\approx 0.2$) 
for $Z=0.040$.   The lack of any
observed Hyades WDs with progenitor masses $\geq 4\msun$
\citep{Weidemann92,Claver01} is
mildly inconsistent for the $\Gamma=2$ power law and Naylor
IMFs at $\mcrit=6\msun$ ($P=0.13$ and $P=0.07$, respectively) and for the
$\Gamma=2$ power-law IMF for $\mcrit=8\msun$ ($P=0.07$), but is
inconsistent with the survey results for the remaining explored
parameter space. This result will be explored in the
discussion (\S4) below.

Of the eight photometrically-detectable Hyades WDs, one is in a binary
system. This is inconsistent with the number of binaries detectable
from a Salpeter IMF ($P\lesssim 0.05$) and mildly inconsistent with
the Kroupa IMF and $\Gamma=2$ power-law IMF
for the binary fraction of 0.5, but is otherwise 
consistent with the calculations for the $Z=0.019$ systems.
For $Z=0.04$, the single observed binary is inconsistent with one
observed binary for each IMF except the Naylor IMF, which is fully
consistent with the observations.

\citet{Bohmvitense95} reports on an IUE survey for WD companions to F
stars in the Hyades, with only one WD-F star system (actually a triple
system with two F stars and a WD) being detected.  In our
calculations, 
we have identified all stars meeting the \bv ~selection criteria in that
study.  The average number of WD companions to these F stars are
presented in the final column of Table \ref{tab.hyades_sim}.
On average, $\lesssim 1$ F star-WD binary is present in a simulated
cluster, fully consistent with the \citet{Bohmvitense95} results.  

\emph{Praesepe} --- \citet{Claver01} list ten candidate
\objectname{Praesepe} 
WDs. Based on their analysis of six spectra, five are cluster WDs, and
the other is a confirmed non-member.  We obtained spectroscopy of
two of the remaining candidates, \objectname{WD 0834+209} and
\objectname{LB 1839}, with the upgraded blue camera of LRIS
\citep{Oke95} on the Keck 10-m telescope in December 2002.  Both of
these objects are found to be QSOs, with  \objectname{LB 1839} at
$z=0.86$ and \objectname{WD 0834+209} at $z=1.09$.
Therefore, at most seven of the Praesepe candidates are WDs, and this
is the number of Praesepe WDs used in the comparison to the calculated
populations.

The realizations of the Praesepe calculations were normalized by the luminosity
function of \citet{Jones91}, who find 179
stars (completeness-corrected) with $2.5\leq M_V \leq 7.5$. 
While the \citet{Jones91}
data extend to fainter magnitudes,   \citet{Claver01} find that the
Padova isochrones do not match the Praesepe MS below this
magnitude, and this mismatch would systematically bias our calculations.

The age and metallicity of the Praesepe are nearly
indistinguishable from that of the Hyades \citep{Claver01}, so we
adopt the Hyades logarithmic cluster age of 8.80 for Praesepe.  
As the metallicity was found to have only
a small effect on the calculated observable number of WDs in the Hyades, 
we simulate Praesepe using only the Z=0.019 isochrones.
The binary star fraction of the Praesepe is $\sim 0.4$
\citep{Mermilliod99}, so this value is used in the simulation.  

For WD detection criteria, we require $\ub\leq 0.0$ and $\bv\leq 0.6$,
limits encompassing the \citet{Claver01} WD sample.  As in the Hyades,
the form of the IMF and \mcrit were varied.  The calculated WD
populations
are given in Table \ref{tab.praesepe_sim}.

As with the Hyades, the observed WD numbers are
inconsistent with the calculated output for a Salpeter and Kroupa
IMFs. The numbers of observable WDs calculated using the 
Naylor law IMF are consistent with the observed number for
$\mcrit=6\msun$ ($P=0.24$) but mildly inconsistent for higher values
of \mcrit ($P=0.15$ and $P=0.11$ for $\mcrit=8\msun$ and
$\mcrit=10\msun$, respectively).  The $\Gamma=2$ power-law IMF
simulation results are fully consistent with the observed numbers.

\citet{Claver01} detected one WD with a progenitor mass $\geq
4\msun$. This number is inconsistent with a Salpeter and Kroupa IMF for all
values of \mcrit ($P\leq 0.04$).  The simulated number of WDs with
massive progenitors is inconsistent with the Naylor IMF for
$\mcrit=10\msun$ ($P=0.05$), but consistent with the one observed WD
for $\mcrit=8\msun$ and $\mcrit=6\msun$ ($P=0.17$ and $P=0.20$,
respectively).  The $\Gamma=2$ power law
WD numbers are reasonably consistent with the observations for all
values of \mcrit.

None of the five confirmed Praesepe WDs are known to be in binary systems.
If neither of the two remaining candidates is in a binary system, then
the lack of binaries is inconsistent with the simulated results for a
Salpeter or Kroupa IMF.  
The lack of binaries is mildly inconsistent  ($P=0.08$)
with the $\Gamma=2$ IMF, and consistent ($P\approx 0.19$) with the
Naylor IMF.  It is emphasized that this inconsistency is
between the lack of observed WD binaries in Praesepe and the number of
simulated WDs in binaries which would be detected given the stated
photometric selection criteria; this discrepancy does \emph{not}
include the WDs in binary systems with photometric indices outside the
selection criteria.

\emph{Pleiades} --- The \objectname{Pleiades} has one known member WD,
\objectname{LB 1497}.   For this cluster, we have normalized the IMF
such that each simulated cluster contains 107 stars in the magnitude
range $2.0\leq V \leq 10.0$, consistent with
the luminosity function of the Pleiades from
\citet{Jones70}.  We assume the \citet{Pinsonneault98} distance
modulus of 5.60 and reddening of $\ebv=0.04$.  The metallicity of the
Pleiades is near solar, with $[\mathrm{Fe}/\mathrm{H}]=-0.034\pm
0.024$ \citep{Boesgaard90}, so we use the Z=0.019 isochrones.
We adopt the \citet{Mermilliod92} multiple-star frequency 
of 0.36 as the binary frequency.  
Age estimates of the Pleiades have shown scatter, with recent
estimates  between 100 Myr
\citep{Pinsonneault98} and 125 Myr \citep{Stauffer98}. We therefore
produce calculations for logarithmic cluster ages of 8.0 and 8.1. 

The results are given in Table
\ref{tab.pleiades_sim}. For $\mcrit=8\msun$ and $\mcrit=10\msun$, the
Salpeter and Kroupa IMFs are inconsistent with the one observed WD,
regardless of 
the adopted cluster age.  For $\mcrit=6\msun$, the Salpeter and Kroupa
IMFs are mildly
inconsistent with the observed WD if the Pleiades have a logarithmic
age of 8.1 ($P=0.12$), but are consistent for a logarithmic age of
8.0.  The $\Gamma=2$ and Naylor IMFs produce WD populations 
mildly inconsistent with the Pleiades for $\mcrit=10\msun$ and for the
$\mcrit=8\msun, \log(\mathrm{age})=8.1$ simulations.  Otherwise,
these two IMFs result in WD populations consistent with the single
observed WD.

\section{Discussion\label{section.discussion}}

 A few observations can be made from the comparisons between the
 calculated WD populations and the observed populations
presented above. The calculations consistently over-predict
the number of observed, color-selected WDs and the number of
color-selected WDs in binary systems. However, the discrepancy between
the expected number of WDs and the observed number is lower than in
previous observational 
studies due to the inclusion of binaries in the simulation.
Therefore, preferential evaporation of WDs from these clusters due
to dynamical evolution need not be as severe as suggested by
 \citet{Weidemann92} but cannot be ruled out.  It should also be noted
 that the binary fractions used in this work represent the high end of
 the range of commonly-quoted cluster binary populations.  A lower
 actual binary fraction would result in a larger WD deficit.

The number of observed WDs can be brought into agreement with the
calculated numbers via at least four different methods.  First, the
slope of the IMF could be steeper than the slopes used here.  Second,
the binary fractions may be higher, resulting in more WDs being hidden
in unresolved binaries.  Third, the binary mass ratio could be closer
to unity than the random parings discussed here, which would result in
more binary WDs being hidden in systems.  Fourth, 
some sort of dynamical evolution may
be removing WDs from the open clusters.  From the simulation alone, it
is not possible to differentiate between these scenarios.  

As mentioned above, the lack of observed WDs in the Hyades and
Praesepe with progenitor masses $\gtrsim 4\msun$ is inconsistent with
the calculated populations.  A steeper high-mass IMF would explain
the lack of these WDs, as would dynamical evolution.  However, there
is some evidence that binary stars in this mass range tend to have
a fairly flat distribution of mass ratios, especially for orbital
periods $\lesssim 0.1{\rm yr}$ \citep{Abt90,Mason98}.  If this is
true, WDs with massive progenitors may be more likely to be in
binary systems with massive, unevolved stars than in binary systems
with fainter, low-mass stars, and therefore more likely to go
undetected.

There are several avenues of research which could shed some
light on these issues.  First, systematic, thorough searches for
WD-mass companions to stars in the Hyades and Praesepe are needed to
complete the census of WDs in each cluster, thereby determining
whether or not binary systems can account for the majority of
``missing'' WDs in open clusters.  Such a survey could include a
combination of spectroscopic and high-resolution imaging (e.g. speckle
and adaptive optics surveys) of these clusters, such as those
published in \citet{Patience98} and similar surveys. These surveys
could also determine if the binary mass ratios for any close binaries
are non-random, as opposed to the random ratios assumed in this work.

Also, WD searches in open clusters of a wide range of ages, such as
the LAWDS survey we are currently undertaking, can provide evidence as to
whether the deficit of massive WDs is increasing with time, as
would be expected for dynamical evolution.  In addition, detailed
comparisons of clusters of similar ages but differing compactness and
richness should also show differences in the white dwarf deficit if
dynamical evolution of the WD population is significant. 

The calculations presented here will be useful in interpreting
data from the WD observational programs.  Most importantly, the
calculations permit us to estimate how strong the white dwarf deficit
is in a cluster, given basic assumptions about the IMF and binary
fraction.  This provides a better estimate on the significance of any
apparent deficit and the cluster-to-cluster variations in the deficit than
could be gleaned from simpler WD population estimators, such as simple
integration of the IMF. 

\acknowledgements
The author wishes to thank Michael Bolte for many helpful comments
regarding this paper and for his skillful advising throughout this
project.  Thanks are also due to Matthew Wood, William Mathews, Graeme
Smith, James Liebert, Brad Hansen and Jasonjot Kalirai for helpful discussions
regarding this work.  Thanks to Chuck Claver for finding charts for
the Praesepe WD candidates.  White dwarf spectral models were
graciously and freely provided by Detlev Koester, and white dwarf
evolutionary models were graciously and freely provided by Matt Wood;
these are greatly appreciated.  The author would also like to thank the 
anonymous referee for insightful comments which led to improvements in this 
manuscript.
Some of the data presented herein were
obtained at the  W.M. Keck Observatory, which is operated as a
scientific partnership among the California Institute of Technology,
the University of California and the National Aeronautics and Space
Administration. The Observatory was made possible by the generous
financial support of the W.M. Keck Foundation. The author wishes to
recognize and acknowledge the very significant cultural role and
reverence that the summit of Mauna Kea has always had within the
indigenous Hawaiian community.  We are most fortunate to have the
opportunity to conduct observations from this mountain.

\clearpage
\begin{deluxetable*}{cccclcccc}
\tablecolumns{9}
\tablecaption{Simulation of the Hyades.\label{tab.hyades_sim}}
\tablewidth{0pt}
\tablehead{\colhead{\mcrit} & \colhead{Z} & \colhead{Binary} & \colhead{IMF\tablenotemark{a}} & &  \colhead{$N_{WD}$} & \colhead{$N_{WD}$} &
\colhead{$N_{WD}$} & \colhead{$N_{WD,F}$\tablenotemark{b}} \\
\colhead{(\msun)} & &\colhead{Fraction} & & & & in binaries & ($M_i>4\msun$) & }
\startdata
8 & 0.019 & 0.4 & S & total    & $29.7$ & $16.9$ & $14.2$ & $0.4$ \\
  &       &     &   & observed & $17.5$ & $4.7$  & $6.7$  & \nodata      \\
  &       &     & P & total    & $14.1$ & $8.0$  & $5.6$  & $0.07$\\
  &       &     &   & observed &  $9.2$ & $3.2$  & $2.7$  & \nodata      \\
  &       &     & N & total    & $18.9$ & $10.7$ & $8.0$  & $0.8$ \\
  &       &     &   & observed &  $9.8$ & $1.7$  & $3.6$  & \nodata      \\
  &       &     & K & total    & $31.9$ & $18.0$ & $15.4$ & $0.8$ \\ 
  &       &     &   & observed & $16.9$ & $3.3$  & $7.1$  & \nodata \\
  &       & 0.5 & S & total    & $29.9$ & $19.7$ & $14.2$ & $0.5$ \\
  &       &     &   & observed & $15.5$ & $5.5$  & $5.4$  & \nodata      \\
  &       &     & P & total    & $14.2$ & $9.5$  & $5.6$  & $0.08$\\
  &       &     &   & observed & $8.4$  & $3.7$  & $2.2$  & \nodata      \\
  &       &     & N & total    & $19.2$ & $12.6$ & $8.1$  & $0.9$ \\
  &       &     &   & observed & $8.4$  & $2.0$  & $2.9$  & \nodata      \\
  &       &     & K & total    & $32.0$ & $21.1$ & $15.5$ & $0.9$ \\
  &       &     &   & observed & $14.5$ & $3.8$  & $5.7$  & \nodata \\
  & 0.040 & 0.4 & S & total    & $32.2$ & $18.2$ & $15.1$ & $0.5$ \\
  &       &     &   & observed & $20.7$ & $7.0$  & $7.6$  & \nodata      \\
  &       &     & P & total    & $15.7$ & $9.0$  & $6.2$  &$0.077$\\
  &       &     &   & observed & $11.3$ & $4.6$  & $3.2$  & \nodata      \\
  &       &     & N & total    & $20.6$ & $11.6$ & $8.6$  & $0.8$ \\
  &       &     &   & observed & $11.3$ & $2.5$   & $4.0$  & \nodata      \\
  &       &     & K & total    & $34.4$ & $19.3$ & $16.4$ & $0.9$ \\
  &       &     &   & observed & $19.6$ & $4.9$  & $7.9$  & \nodata \\
6 & 0.019 & 0.4 & S & total    & $25.4$ & $14.4$ & $9.9$  & $0.4$ \\
  &       &     &   & observed & $15.4$ & $4.5$  & $4.7$  & \nodata      \\
  &       &     & P & total    & $12.7$ & $7.2$  & $4.2$  &$0.064$\\
  &       &     &   & observed &  $8.5$ & $3.1$  & $2.0$  & \nodata      \\
  &       &     & N & total    & $16.8$ & $9.4$  & $5.8$  & $0.7$ \\
  &       &     &   & observed &  $8.9$ & $1.7$  & $2.7$  & \nodata      \\
  &       &     & K & total    & $27.1$ & $15.3$ & $10.6$ & $0.7$ \\
  &       &     &   & observed & $14.7$ & $3.1$  & $5.0$  & \nodata \\
10& 0.019 & 0.4 & S & total    & $32.2$ & $18.2$ & $16.6$ & $0.5$ \\
  &       &     &   & observed & $18.6$ & $4.8$  & $7.7$  & \nodata      \\
  &       &     & P & total    & $14.9$ & $8.5$  & $6.3$  &$0.076$\\
  &       &     &   & observed &  $9.6$ & $3.2$  & $3.0$  & \nodata      \\
  &       &     & N & total    & $20.0$ & $11.3$  & $9.1$  & $0.87$ \\
  &       &     &   & observed & $10.3$ & $1.7$  & $4.1$  & \nodata      \\
  &       &     & K & total    & $34.5$ & $19.4$ & $18.1$ & $0.8$ \\
  &       &     &   & observed & $18.1$ & $3.3$  & $8.3$  & \nodata \\
\enddata
\tablenotetext{a}{S=Salpeter, $\Gamma=1.35$; P=power law, $\Gamma=2$; N=Naylor IMF, $\Gamma=0.2 (M\leq 1\msun)\,,\,\Gamma=1.8 (M>1\msun)$; K=Kroupa IMF,$\Gamma=0.3 (0.08\sun\geq M> 0.5\msun),\,\Gamma=1.3 (M\geq 0.5\msun)$}
\tablenotetext{b}{Number of WDs in binaries with F stars (see text)}
\end{deluxetable*}

\begin{deluxetable*}{cccccccc}
\tablecolumns{8}
\tablecaption{Simulation of the Praesepe.\label{tab.praesepe_sim}}
\tablewidth{0pt}
\tablehead{\colhead{\mcrit} & \colhead{IMF\tablenotemark{a}} & 
  \colhead{$N_{WD}$} & \colhead{$N_{WD}$} & \colhead{$N_{WD}$}& 
  \colhead{$N_{WD}$} & \colhead{$N_{WD}$} & \colhead{$N_{WD}$} \\
\colhead{(\msun)} & & \colhead{total} & \colhead{observed}&
\colhead{total} & \colhead{observed}& \colhead{total} & 
\colhead{observed} \\  
&&&&\colhead{in binaries} & \colhead{in binaries} &
\colhead{($M_i>4\msun$)}& \colhead{($M_i>4\msun$)} }
\startdata
10& S & $35.1$ & $19.7$ & $19.8$  & $4.8$ & $18.0$ & $8.4$\\
  & P & $13.2$ & $ 8.1$ & $ 7.6$  & $ 2.6$ & $5.6$ & $2.7$ \\
  & N & $23.0$ & $11.5$ & $12.9 $ & $ 1.7$ & $10.4$ & $4.7$ \\
  & K & $38.7$ & $20.3$ & $21.8$  & $ 3.7$ & $20.2$ & $9.3$ \\
8 & S & $32.5$ & $18.5$ & $18.4$  & $ 4.7$ & $15.5$ & $7.3$ \\
  & P & $12.6$ & $7.9$ & $7.2$ & $2.6$ & $5.0$ & $2.4$\\
  & N & $21.8$ & $10.9$ & $12.3$ & $1.7$ & $9.3$ & $3.2$\\
  & K & $35.7$ & $18.9$ & $20.1$ & $3.6$ & $17.3$ & $8.0$ \\
6 & S & $27.8$ & $16.3$ & $15.8$ & $4.5$ & $10.7$ & $5.1$ \\
  & P & $11.3$ & $7.3$ & $6.5$ & $2.5$ & $3.7$ & $1.9$\\
  & N & $19.2$ & $9.8 $ & $10.8$ & $1.6$ & $6.7$ & $3.0$\\
  & K & $30.3$ & $16.5$ & $17.1$ & $3.5$ & $11.9$ & $5.6$ \\
\enddata
\tablenotetext{a}{S=Salpeter, $\Gamma=1.35$; P=power law, $\Gamma=2$; N=Naylor IMF, $\Gamma=0.2 (M\leq 1\msun)\,,\,\Gamma=1.8 (M>1\msun)$; K=Kroupa IMF,$\Gamma=0.3 (0.08\sun\geq M> 0.5\msun),\,\Gamma=1.3 (M\geq 0.5\msun)$}
\end{deluxetable*}

\begin{deluxetable*}{ccccccc}
\tablecolumns{7}
\tablecaption{Simulation of the Pleiades.\label{tab.pleiades_sim}}
\tablewidth{0pt}
\tablehead{\colhead{\mcrit} & \colhead{log(age)} & \colhead{IMF\tablenotemark{a}} & 
  \colhead{$N_{WD}$} & \colhead{$N_{WD}$} & \colhead{$N_{WD}$}& 
  \colhead{$N_{WD}$}  \\
\colhead{(\msun)} && & \colhead{total} & \colhead{observed}&
\colhead{in binaries} & \colhead{observed in binaries}} 
\startdata
10& 8.0 & S & $9.1$ & $7.6$ & $4.8$ & $3.3$ \\
  &     & P & $3.9$ & $3.5$ & $2.1$ & $1.7$ \\
  &     & N & $5.2$ & $3.6$ & $2.7$ & $1.1$ \\
  &     & K & $9.8$ & $7.2$ & $5.1$ & $2.6$ \\
  & 8.1 & S & $11.3$& $9.1$ & $6.0$ & $3.9$ \\
  &     & P & $4.9$ & $4.4$ & $2.6$ & $2.0$ \\
  &     & N & $6.6$ & $4.4$ & $3.5$ & $1.3$ \\ 
  &     & K & $12.1$& $8.6$ & $6.4$ & $2.9$ \\
8 & 8.0 & S & $6.7$ & $5.7$ & $3.6$ & $2.5$ \\
  &     & P & $2.9$ & $2.7$ & $1.6$ & $1.3$ \\  
  &     & N & $3.9$ & $2.8$ & $2.0$ & $0.9$ \\
  &     & K & $7.2$ & $5.3$ & $3.8$ & $2.0$ \\
  & 8.1 & S & $8.9$ & $7.3$ & $4.7$ & $3.2$ \\
  &     & P & $4.1$ & $3.6$ & $2.2$ & $1.8$ \\
  &     & N & $5.3$ & $3.6$ & $2.8$ & $1.1$ \\
  &     & K & $9.5$ & $6.9$ & $5.0$ & $2.4$ \\
6 & 8.0 & S & $2.3$ & $2.0$ & $1.2$ & $0.9$ \\
  &     & P & $1.1$ & $1.0$ & $0.6$ & $0.5$ \\
  &     & N & $1.4$ & $1.0$ & $0.7$ & $0.4$ \\
  &     & K & $2.4$ & $1.9$ & $1.3$ & $0.8$ \\
  & 8.1 & S & $4.4$ & $3.7$ & $2.3$ & $1.7$ \\
  &     & P & $2.1$ & $2.0$ & $1.1$ & $1.0$ \\
  &     & N & $2.8$ & $2.0$ & $1.5$ & $0.7$ \\
  &     & K & $4.7$ & $3.5$ & $2.5$ & $1.3$ \\
\enddata
\tablenotetext{a}{S=Salpeter, $\Gamma=1.35$; P=power law, $\Gamma=2$; N=Naylor IMF, $\Gamma=0.2 (M\leq 1\msun)\,,\,\Gamma=1.8 (M>1\msun)$; K=Kroupa IMF,$\Gamma=0.3 (0.08\sun\geq M> 0.5\msun),\,\Gamma=1.3 (M\geq 0.5\msun)$}
\end{deluxetable*}

\end{document}